\documentclass[preprint,prd,letterpaper,tightenlines,nofootinbib,showpacs]
{revtex4}
\usepackage{amsmath,amssymb,graphicx,mathrsfs}
\usepackage{dcolumn}
\usepackage{bm}
\def\sfrac#1#2{{\textstyle \frac{#1}{#2}}}
\newcommand{\nn}{\nonumber\\}

\newcommand {\boldgamma}{\mbox{\boldmath$\gamma$}}

\def\bfn {{\bf n}}

\def\bfK{{\bf K}}
\newcommand{\bb}{\langle}
\newcommand{\kk}{\rangle}

\def\bfr{{\bf r}} \def\bfs{{\bf s}}

\newcommand{\ben}{\begin{displaymath}}
\newcommand{\een}{\end{displaymath}}
\newcommand{\be}{\begin{equation}}
\newcommand{\ee}{\end{equation}}
\newcommand{\bea}{\begin{eqnarray}}
\newcommand{\eea}{\end{eqnarray}}
\newcommand{\eqn}[1]{\label{#1}}
\newcommand{\eq}[1]{Eq.~(\ref{#1})}

\begin{document}
\vspace{1.0cm}
\vspace{.50cm}
\title{\begin{flushright}{\normalsize NT@UW-07-01}\end{flushright}
Subtleties  of Lorentz Invariance and shapes of the 
Nucleon}

\author{Alexander  Kvinikhidze}
\affiliation{The Mathematical Institute of 
Georgian Academy of Sciences, Tbilisi, Georgia} 
\author{Gerald A. Miller}
\affiliation{ University of Washington
  Seattle, WA 98195-1560}

\sloppy

\begin{abstract}

We study the effects of 
 Lorentz invariance
on 
relativistic constituent quark model 
wave functions.
The  
model nucleon wave function of Gross {\it et al.} is  constructed such 
that there is no orbital angular momentum and that the spin-dependent density 
is spherical. This model wave function is  
  claimed to be manifestly  covariant. 
We consider two possible interpretations of the nucleon wave function in an 
arbitrary reference frame.  
  In the first, 
the seeming covariance of the matrix elements of the
electromagnetic (em) current 
arises from  using   the Breit frame. 
Matrix elements have a different appearance in
any other frame. In the second interpretation, the em current is covariant yet it is not consistent 
with the general structure required by QFT, e.g. 
the wave function of the incoming nucleon depends on the momentum of the outgoing nucleon and vice versa.

\end{abstract}
\maketitle
\vskip0.5cm

\section{Introduction}

Recent 
Jefferson Laboratory  data on em form factors of the nucleon
have created much theoretical interest. 
The key finding is  that the ratio of the proton's
$G_E/G_M$ falls rapidly with increasing $Q^2$\cite{Jones99,Gayou01}. 
But new results for
the neutron electric and magnetic form factors have been 
or are about to be obtained; see the reviews 
\cite{Hyde-Wright:2004gh,Gao:2005cj,Perdrisat:2006hj}. 

It was 
 argued\cite{Miller:2002qb}
 that  reproducing the measured
ratio $G_E/G_M$
ratio requires a relativistic treatment that includes
the  effects of the quark's
non-zero  orbital
angular momentum. Ref.~\cite{me} introduced 
the idea of using the rest-frame to rest-frame matrix elements of 
a spin-dependent charge density operator
to exhibit the influence of the orbital angular momentum.
In particular, (for a model without explicit gluons) 
the probability for a quark to  have a given momentum, $\bfK$, 
 and a given direction
of spin, $\bfn$ is given by  \cite{me,us}
\bea\widehat{\rho}_{\cal O}(\bfK,\bfn)=\int \frac{d^3r}{(2\pi)^3} e^{i\bfK\cdot\bfr}
\bar{\psi}(\bfr){\cal O}\frac{1}{2}
(\gamma^0+\boldgamma\cdot\bfn\gamma_5)\psi({\bf 0}), 
\label{qft}\eea 
where 
 $\cal O$ is  
$\widehat{Q}/e$, the quark charge operator in units of the proton charge for the 
spin-dependent charge density or 
${\cal O}=1$ for the  spin-dependent matter density.
The  matrix element of the operator $\widehat{\rho}_{\cal O}(\bfK,\bfn)$
 gives the spin-dependent matter densities.
The quark  field
operators $\bar{\psi}(\bfr), \bar{\psi}({\bf 0})$ 
are evaluated at equal time.   The rest-frame matrix 
element of this density operator in
a nucleon state of definite total angular momentum  defined
 by the unit vector $\bfs,\;\vert\Psi_{\bfs}\kk$  is
\bea
\rho_{\cal O}(\bfK,\bfn,\bfs)
\equiv\bb\Psi_\bfs\vert \widehat{\rho}_{\cal O}(\bfK,\bfn)\vert\Psi_{\bfs}\kk
,\label{sdd}\eea
where the subscript ${\cal O} =Q,$ or ${\cal O}=\;1$ specifies the operator
 used in \eq{qft}.
The most general shape of the proton, 
obtained if parity and rotational invariance are upheld can be written as
\bea
\rho_{\cal O}(\bfK,\bfn,\bfs)=A_{\cal O}(\bfK^2)+B_{\cal O}(\bfK^2)\bfn\cdot\bfs+C_{\cal O}(\bfK^2)
\left(\bfn\cdot\bfK\;\bfs\cdot\bfK-\frac{1}{3}\bfn\cdot\bfK\; \bfK^2\right)  \label{genshape}
,\eea
with the last term generating the non-spherical shape. 
The effects of non-vanishing orbital angular momentum cause the matrix elements
of the spin-dependent density operator \eq{sdd} to be non-spherical and 
 yield a non-zero value of the coefficient $C_{\cal O}(\bfK^2)$. 
While no experiment has been constructed
to measure the spin-dependent density, this quantity can be evaluated using
the techniques of Lattice QCD, 
and has been measured in condensed matter physics\cite{Prokes}.
 
Gross \& Agbakpe\cite{Gross06} constructed a relativistic constituent quark
model that was claimed to have  a  spherical 
shape. However, these authors did not consider the spin-dependent density operator.
When  we \cite{us} used the wave function of \cite{Gross06}
to evaluate the matrix element of the spin-dependent charge and matter 
density operators, a non-spherical
nucleon shape was obtained. 
More recently,
Gross {\it et al.}  \cite{ramalho} claimed to find a covariant constituent quark-diquark 
model that describes
 all the available em form factors,  but has no
orbital angular momentum. 
In this case, the shape of the proton as determined by
the rest-frame 
matrix element of the spin-dependent
density matrix is indeed spherical.
 The question of whether or not it is possible to find
a covariant model that is pure S-wave is an interesting one that
 we examine 
 here. 

We next  describe 
 the  wave function  of  \cite{ramalho}, using the notation of that reference.
 The nucleon wave function $\Psi_N(P,k)$, 
of total four-momentum $P$ and di-quark four-momentum $k$, is given by 
the expression
\bea\label{psiRel}
\Psi_N (P,k)&=&
\frac{1}{\sqrt{2}} \psi_0(P,k) \, \phi^0_I \, u({\bf P},s) 
 \nn
& &-\frac{1}{\sqrt{6}} \psi_1(P,k) \; 
\phi^1_I \, 
 \gamma_5 \not\!\varepsilon^*_P \,u({\bf P},s),  
\eea
which is a sum of
 contributions from a spin-isospin (0,0) diquark and a spin-isospin (1,1) diquark and 
$\psi_{0,1}$ are Lorentz scalar functions.  
The polarization vectors $\varepsilon_P$ are given by the expression

\bea
\varepsilon_P= {\cal O}_P \epsilon_k,
\label{eta_relation}
\eea 
where $\epsilon_k$ is a genuine 
relativistic polarization vector of a vector particle
 (di-quark in the present case) $\epsilon_k\cdot k=0$. 
This quantity is denoted by $\eta=\epsilon_k$ 
in \cite{ramalho}.
The operator  ${\cal O}_P$  is a 
Lorentz transformation, with
\bea
{\cal O}_P = B_P B^{-1}_k R^{-1}_{\hat k} . \label{gboost}
\eea
The 
operator $R^{-1}_{\hat k}$ rotates ${\bf k}$ 
from a generic $(\theta,\varphi)$ direction to the 
positive $z$ direction.   
$B^{-1}_k$ boosts 
the four-momentum state  $(E_s,0,0,\mbox{k})$ to the diquark rest frame $(m_s,0,0,0)$, 
and finally $B_P$ boosts the vector $(M,0,0,0)$ to the moving frame
$(E_P,0,0,\mbox{P})$. The wave function $\Psi_N$ satisfies the Dirac equation
because $\not\!P$ commutes 
with $\gamma_5\!\!\! \not\!\varepsilon^*_P$. As stressed in ref.~\cite{ramalho},
the essential difference between this model 
and the one introduced in Ref.~\cite{Gross06} is that 
in the nucleon rest frame, the wave function 
(\ref{psiRel}) contains absolutely {\it no\/} 
angular dependence of any kind.  

We discuss the general requirements for covariance  
and a proper treatment of a relativistic constituent (quark di-quark) model. 
We  study two interpretations  of \cite{ramalho} based on two different generalizations of the boost
$B_P$ in \eq{psiRel} to the case of the arbitrary $P$
and find that, using the first (conventional) interpretation, 
the model wave function of \cite{ramalho} does not satisfy these 
requirements because it is not covariant, and as a result produces an
 em form factor which is not Lorentz invariant. Using
 the second (unconventional) interpretation leads indeed to 
a 
 Lorentz invariant em form factor.  
However,
the use of the  second interpretation does not yield a  model that  satisfies
the  basic requirements 
(that any  composite quark model must satisfy)
discussed in  Sect.~\ref{sec:uncon}.
 We summarize in  Sect.~\ref{sec:assess}.

\section{Covariant vector di-quark wave function}
\label{sec:cov}
Let us denote the vector di-quark wave  function as $\Psi_{P,s}(k,\epsilon)$, 
defined as
\bea
\bar{\Psi}_{P,s}(k,\epsilon)=\langle P,s|\bar q(0)|k,\epsilon\rangle,\label{psi1}
\eea
where $\langle P,s|$ and $|k,\epsilon\rangle$ are nucleon and di-quark 
eigenstates  and $\bar q(0)$ represents a 
quantized quark field operator. Note that the dependence on the polarization vector
$\epsilon$ and nucleon spin is made explicit.
 Lorentz invariance requires that 
\bea 
\bar\Psi_{P,s}(k,\epsilon)\sim \bar U(P,s)\Gamma_\mu(P,k)\epsilon_k^\mu,\label{psi2}
\eea 
where $\Gamma_\mu(P,k)$
is a {\it covariant} vector :
\bea
\Gamma_\mu(P,k)=A\gamma_5\gamma_\mu+B\gamma_5k_\mu+C\gamma_5P_\mu+\cdots,\label{gamma}
\eea 
where $A,B,$ and $C$ are Lorentz scalar functions built from the four vectors
$P$ and $k$.  
The forms \eq{psi2}, and \eq{gamma} have been known for a long time \cite{Henriques:1975uh}
and have been applied recently \cite{Bender:2002as,Tiburzi:2004mh}.

To see 
how this Lorentz invariance of \eq{psi2} works in practice, consider the
relevant  particular example of the matrix element of   
 the em 
 current: \cite{diqe} 
\bea\label{corr-f-f}
&&<P_+,s'|J^\alpha(0)|P_-,s>={\cal M}^\alpha=\int d^4k\;\sum_\epsilon\bar\Psi_{P_+,s'}(k,\epsilon)\gamma^\alpha\Psi_{P_-,s}(k,\epsilon)
\nonumber\\&&\sim\int d^4k\;\cdots\sum_\epsilon \bar U(P_+,s')
\Gamma_\mu(P_+,k)\epsilon_k^\mu\gamma^\alpha \epsilon_k^\nu \Gamma_\nu(P_-,k)U(P_-,s),
\eea
where initial and final nucleon four-momentum are denoted as $P_{-}$ and $P_+$.
In Ref.~\cite{ramalho} (their eq.~(11)) $P_{\pm}$ 
are explicitly chosen in the Breit frame. Here the only restriction is that $P_+=P_-+q$
and $P_\pm^2=M^2$, where $q$ is the four-momentum of the virtual photon and
$M$ is the nucleon mass. 
 The quantity ${\cal M}^\alpha$  
 should be explicitly Lorentz invariant.
The sum over polarizations is performed as 
\be
\sum_\epsilon\epsilon_k^\mu \epsilon_k^\nu =\frac{k^\mu k^\nu}{m^2}-g^{\mu\nu}\label{relation}
,\ee
where $m$ is the di-quark mass. Thus        one finds
\be \label{res1}
{\cal M}^\alpha\sim 
\int d^4k\;\cdots\bar U(P_+,s')\Gamma_\mu(P_+,k)\gamma^\alpha(\frac{k^\mu k^\nu}{m^2}-g^{\mu\nu})
\Gamma_\nu(P_-,k)U(P_-,s).      
\ee 
 The result \eq{res1} has a manifestly covariant form as a Lorentz four-vector
that results from the use of
\eq{relation}.

\section{Non-covariant wave function of  \cite{ramalho}}
\label{sec:noncov}
 
 We need the nucleon wave function \eq{psiRel} in an
 arbitrary reference frame. For this we  
use first the straightforward (conventional) definition of the boost  ${\cal L}_P$ which is explicitly stated in \cite{ramalho}, between
 eqs (6) and (7): ``The spin states are analogous of (2) and (3), and their form in an arbitrary frame is obtained by boosting the nucleon to momentum $|{\bf P}|=P$ along the z direction and then rotating". This is equivalent to the statement that the z-direction cannot be favored among others in  \eq{psiRel}. We will show that 
 such model corresponding to the ``first interpretation" of \cite{ramalho}
is not consistent with Lorentz invariance. 
We note in advance that  the 
essential point will be  that  different 
 polarization vectors $\varepsilon_{P_+}$
and  $\varepsilon_{P_-}$ enter into the sum over polarization vectors.

Here consider the following (first interpretation) generalization of the 
vector-diquark part of the nucleon wave function \eq{psiRel} to an
arbitrary reference frame:  
\bea &&
\bar\Psi_P(k)\sim \bar U(P,s)\gamma_5\gamma_\mu\varepsilon_P^\mu =
\bar U(P,s)\gamma_5\gamma_\mu({\cal L}_P{\cal L}^{-1}_k\epsilon_k)^\mu\nonumber\\
&&{\cal L}^{-1}_k\epsilon_k=(0,\vec\epsilon)\equiv \epsilon_0
\eqn{ram}
\eea
 where P and k are the nucleon and diquark on mass shell momenta, 
${\cal L}^{-1}_k$ is the boost transformation ${\cal L}^{-1}_kk=(\sqrt{k^2},{\bf 0})$, 
$\epsilon_k$ is a genuine relativistic polarization vector of a vector particle
 (diquark in the present case) $\epsilon_k\cdot k=0$, 
which is denoted by $\eta=\epsilon_k$ 
in \cite{ramalho}. Furthermore, 
  $\epsilon_0$ 
is the diquark polarization four-vector in the diquark rest frame.
Our notation here
differs slightly from that of \cite{ramalho} because (we use the notation
${\cal L}_k$ instead of 
$R_k B_k$) and because
the quantity ${\cal L}^{-1}_k$ is not exactly the same  as 
$B_k^{-1}R_k^{-1}$ of \eq{gboost}: for the sake of simplicity 
 we not make 
the effects of the rotation explicit. 
This simplification does not affect our conclusions \cite{explain}. 
 
Let us emphasize that the first interpretation consists in replacing 
the boost $B_P$ (with z directed $P$) in the \eq{psiRel} by the boost ${\cal L}_P$ (with arbitrary $P$).
As noted above in \eq{gamma}, Lorentz invariance requires $\Gamma_\mu(P,k)$ 
to be a covariant four-vector 
in any quark-diquark 
wave function $\bar\Psi_P(k)\sim \bar U(P,s)\Gamma_\mu(P,k)\epsilon_k^\mu$.
The result \eq{ram} is 
 not consistent with this requirement  because the quantity
 $\gamma_5\gamma_\nu\left({\cal L}_P{{\cal L}^{-1}_k}\right)^{\nu}_\mu$ 
is not a four-vector.
In particular, the explicit appearance of the product
of boosts, ${\cal L}_P,{\cal L}^{-1}_k$, breaks covariance.
Neither   ${\cal L}_P$ nor ${\cal L}^{-1}_k$ is a
 covariant tensor. To see this we derive
the boost tensor from the expression for the boost Eq.~(2.8) of
\cite{khel} (note that  a sign misprint in that equation is fixed here): 
\bea
({\cal L}^{-1}_k)^\mu_\nu&=&\delta^\mu_\nu
-\frac{1}{(k^0+m)m}k_\nu k^\mu
-\frac{1}{k^0+m}k_\nu\delta_0^\mu
+\frac{2k^0+m}{(k^0+m)m}\delta_{\nu0} k^\mu
-\frac{m}{k^0+m}\delta_{\nu0} \delta_0^\mu.
\label{boost1}\eea
The result  \eq{boost1}, with the explicit presence of the index 0, makes it clear
that 
$({\cal L}^{-1}_k)^\mu_\nu$ is not a covariant tensor.
Similarly
\bea
({\cal L}_P)^\mu_\nu&=&\delta^\mu_\nu
-\frac{1}{(P^0+M)M}(2\delta_\nu^0P^0-P_\nu)(2\delta^\mu_0P^0- P^\mu)
-\frac{1}{P^0+M}(2\delta_\nu^0P^0-P_\nu)\delta_0^\mu\nn
&&
+\frac{2P^0+M}{(P^0+M)M}\delta_{\nu0} (2\delta^{\mu_0}P^0- P^\mu)
-\frac{M}{P^0+M}\delta_{\nu0} \delta_0^\mu,
\label{boost}\eea
where we have used $P_\mu=2\delta^\mu_0P^0-P^\mu$.

Lorentz invariance is lost if 
one uses the  wave function of \cite{ramalho} 
$\bar\Psi_P(k)\sim \bar U(P,s)\Gamma_\mu(P,k)\epsilon_P^\mu$ because 
the sum over diquark polarization,
$\sum_\epsilon\epsilon_{P_+}^\mu \epsilon_{P_-}^\nu$, that enters in the matrix element of 
 the em  current is not Lorentz invariant.
 Let us calculate 
the diquark polarization sum $D^{\mu\nu}$. We find 
\bea\label{Dmunu1}
&&
D_{\mu\nu}=\sum_\epsilon(\epsilon_{P_+})_\mu (\epsilon_{P_-})_\nu =
\sum_{\alpha\beta}({\cal L}_{P_+})_\mu^\alpha\epsilon_{0\alpha}({\cal L}_{P_-})_\nu^\beta
\epsilon_{0\beta}=
\sum_{i=1,2,3}({\cal L}_{P_+})_\mu^i({\cal L}_{P_-})_\nu^i
.\label{fact}\eea Note that here the quantity 
 $D_{\mu\nu}(P^+,P^-)$ is seen to be a sum of
product functions, with one function depending only on $P^+$ and the other 
depending only on $P^-$. 

We use the expression for the boost \eq{boost} to evaluate \eq{Dmunu1}, with the
result
\bea\label{Dmunu2}
&&
D_{\mu\nu}=
\delta_{\mu0}\delta_{\nu0}\frac{M^2-P_+\cdot P_-}{(P_+^0+M)(P_-^0+M)}
-g_{\mu\nu}
\nn
&&
+\frac{P^+_\mu P^+_\nu}{(P_+^0+M)M}
+\frac{P^-_\nu P^-_\mu}{(P_-^0+M)M}
+\frac{P^+_\mu P^-_\nu {\bf P}_+\cdot{\bf P}_-}{(P_+^0+M)(P_-^0+M)M^2}
\nn
&&
+\frac{\delta_\mu^0P^+_\nu}{P_+^0+M}+\frac{P^-_\mu\delta_\nu^0}{P_-^0+M}
-\frac{P^+_\mu\delta_{\nu0}(P_+P_-+P_+^0M)+\delta_\mu^0P^-_\nu(P_+P_-+P_-^0M)}
{(P_+^0+M)(P_-^0+M)M}.\label{fg}
\eea
A brief inspection shows that  $D_{\mu\nu}$, as obtained in a general reference
frame, involves the non-covariant expressions $\delta_{\mu0}$ as well as explicit
three-vectors and therefore  is not a covariant tensor.
This result means that the wave function of  \cite{ramalho} is  not covariant and that
the expressions
for matrix elements of the
em current that result from using \eq{fg} are not covariant.

However, one can be fooled by using one particular frame--
the Breit frame. In this case, 
the four-vectors $P^\mu_\pm$ are given by 
 $P_+=(E,0,0,Q/2)$, $P_+=(E,0,0,-Q/2)$.  It is also useful to note
that  the non-covariant expression  $\delta^\mu_0=(1,0,0,0)$,
can be written in an apparently covariant form
\be
\delta^\mu_0=\frac{(P_++P_-)^\mu}{\sqrt{4M^2-(P_+-P_-)^2}}\label{delta0}
\ee
We proceed by evaluating \eq{Dmunu2} in  
 the Breit frame.
Use \eq{delta0} and 
${\bf P}_+=-{\bf P}_-={\bf P}$, \\
$P_+\cdot P_-=2P_0^2-M^2$ in \eq{Dmunu2} to obtain
\bea
&&D_{\mu\nu}=(P^+_\mu +P^-_\mu)(P^+_\nu +P^-_\nu)\frac{2M^2-2P_0^2}{4(P^0+M)^2P_0^2}
-g_{\mu\nu}
\nn
&&
+\frac{P^+_\mu P^+_\nu}{(P^0+M)M}
+\frac{P^-_\nu P^-_\mu}{(P^0+M)M}
+\frac{P^+_\mu P^-_\nu (M^2-P_0^2)}{(P^0+M)^2M^2}
\nn
&&
+\frac{(P^+_\mu +P^-_\mu)P^+_\nu}{2P^0(P^0+M)}+
\frac{P^-_\mu(P^+_\nu +P^-_\nu)}{2P^0(P^0+M)}
\nn
&&
-\frac{[P^+_\mu(P^+_\nu +P^-_\nu)+(P^+_\mu +P^-_\mu)P^-_\nu]
[2P_0^2-M^2+P^0M=(P^0+M)(2P^0-M)]}
{2(P^0+M)^2P^0M}
\nn
&&=(P^+_\mu +P^-_\mu)(P^+_\nu +P^-_\nu)\frac{1}{2P_0^2}
-g_{\mu\nu}
-\frac{P^+_\mu P^-_\nu}{M^2}\nn
&&=(P^+_\mu +P^-_\mu)(P^+_\nu +P^-_\nu)\frac{1}{M^2+P_+\cdot P_-}
-g_{\mu\nu}
-\frac{P^+_\mu P^-_\nu}{M^2}
.\label{app}\eea

This result, obtained previously in Ref.~\cite{ramalho},
has a illusory covariant appearance, resulting from the explicit use of the
Breit frame. 
The expression \eq{app} would not be correct in any frame other than 
the 
frame where 3D parts of $P_+$ and $ P_-$ are collinear. 
In particular, the factor $M^2+P_+\cdot P_-$ that appears in 
the denominator of \eq{app} violates the
sum of product functions form of  \eq{fact}.

\section{ Wave function with  unconventional polarization vectors}
\label{sec:uncon}

We found out above that the first interpretation  wave function of \cite{ramalho} is not covariant, and as a result it fails to produce 
a 
Lorentz invariant em form factor. Then we can ask the question, what the wave function of \cite{ramalho} should be in the arbitrary reference frame to produce the Lorentz invariant em form factor
given that the z direction moving nucleon is described by the wave function of \eq{psiRel}?
The unambiguous solution to this question is to use the ``unconventional" diquark polarization vectors $\xi(P_\pm)$
\footnote{We understand that 
these  polarization vectors are suggested (and named ''unconventional") 
by F. Gross \cite{Franz} for the construction of the pure S-wave wave function.} 
in the covariant wave function (\ref{psi2}) and correspondingly in \eq{corr-f-f} instead of usual $\epsilon_k$, 
 \be
\xi(P_+) =\Lambda\epsilon(Z_+) =\Lambda{\cal L}_{Z_+} \epsilon_0,
\hspace{1cm}\xi^T(P_-)= \epsilon^T(Z_-)\Lambda^T = \epsilon^T_0{\cal L}^T_{Z_-}\Lambda^T, \eqn{unconv}
 \ee
 where  $\Lambda={\cal L}_{P_++P_-}$ is the boost transformation defined as 
 \be
 \Lambda^{-1}(P_++P_-)=(\sqrt{(P_++P_-)^2},{\bf 0})=Z_++Z_-,\hspace{1cm} Z_\pm=\Lambda^{-1} P_\pm.
 \ee
  So the 4-vectors $Z_+$ and $Z_-$ are collinear, $\vec Z_-=-\vec Z_+$. Indeed Lorentz invariance of the em current implies that it transforms under the Lorentz transformation $L$ as \footnote{Just the same arguments of Lorentz invariance for the quark di-quark wave function (\ref{psi1}) unambiguously lead to the {\it covariant} form 
  (\ref{psi2}). This paper addresses possible violations of general principles if 
 forms inconsistent with  (\ref{psi2}) 
  are used.}
  \bea\label{boos-ff}
&&<P_+,s'|J^\alpha(0)|P_-,s>= {\cal M}^\alpha=\bar U(P_+,s')M^\alpha
(P_+,P_-)U(P_-,s),\nn
&&
M^\alpha (LP_+,LP_-)=L^\alpha_\beta S_L  M^\beta(P_+,P_-)S^+_L,\hspace{5mm}
\mbox{where}\hspace{5mm}
S_L \gamma^\alpha S^+_L=(L^{-1})^\alpha_\beta\gamma^\beta
\eea
Therefore the em current can be boosted to the arbitrary frame as follows
\be
M^\alpha(P_+,P_-)=M^\alpha (\Lambda Z_+,\Lambda Z_-)=
\Lambda^\alpha_\beta S_\Lambda M^\beta(Z_+,Z_-)S^+_\Lambda,
\ee
where
\be
M^\alpha(Z_+,Z_-)\sim\sum_i\int d^4k\delta^+(k^2-m^2)\psi_1(Z_+,k)
\epsilon_i^\mu(Z_+)\epsilon_i^\nu(Z_-)\gamma_\mu\gamma_5\gamma^\alpha\gamma_5\gamma_\nu\psi_1(Z_-,k)
\ee
is expressed in terms of the wave functions, (\ref{psiRel}), of nucleons moving along z direction.
 Simple algebra leads unambiguously to the unconventional polarization vectors of \eq{unconv}:
\bea\label{M-apha}
&&
M^\alpha(P_+,P_-)\sim\Lambda^\alpha_\beta\sum_i\int d^4k\delta^+(k^2-m^2)\psi_1(Z_+,k)
\epsilon_i^\mu(Z_+)\epsilon_i^\nu(Z_-)S_\Lambda \gamma_\mu\gamma^\beta\gamma_\nu S^+_\Lambda\psi_1(Z_-,k)
\nn
&&
=\sum_i\int d^4k\delta^+(k^2-m^2)\psi_1(P_+,\Lambda k)
[\Lambda\epsilon_i(Z_+)]^\mu[\Lambda\epsilon_i(Z_-)]^\nu \gamma_\mu\gamma^\alpha\gamma_\nu 
\psi_1(P_-,\Lambda k)
\nn
&&
=\sum_i\int d^4k\delta^+(k^2-m^2)\psi_1(P_+, k)
\xi_i^\mu(P_+)\xi^\nu_i(P_-) \gamma_\mu\gamma^\alpha\gamma_\nu 
\psi_1(P_-, k),
\eea
where Lorentz invariance for the scalar functions was used, $\psi_1(Z_\pm,k)=\psi_1(P_\pm, \Lambda k)$. Now,
using (\ref{M-apha}) in the first line of the \eq{boos-ff}, the expression for the em form factor in terms of the new (unconventional) polarization vectors reads 
\bea\label{resp}
&&{\cal M}^\alpha\sim \sum_i\int d^4k\delta^+(k^2-m^2)\bar U(P_+)\bar\Gamma'_\mu(P_+,k)
\xi^\mu(P_+,i)\bar\xi^\nu(P_-,i)\gamma^\alpha\Gamma'_\nu(P_-,k)U(P_-)
\nn
&&
=\int d^4k\delta^+(k^2-m^2)\bar U(P_+)\bar\Gamma'_\mu(P_+,k)
D^{\mu\nu}(P_+,P_-)\gamma^\alpha\Gamma'_\nu(P_-,k)U(P_-)
\eea
where $\Gamma'_\nu(P_-,k)=\gamma_5\gamma_\nu\psi_1(P_-,k)$ corresponds to the wave function (\ref{psiRel}). Below the primed $\Gamma'_\nu(P_-,k)$ denote more general matrices used in the \eq{gen-r-wf}, i.e.  in the wave functions of \cite{ramalho} corresponding to the round spin-dependent density. They should not be confused with the (unprimed) $\Gamma_\nu(P_-,k)$ used in the covariant wave functions (\ref{psi2}), which according \eq{gamma} are strictly covariant and do not depend on the momentum of other nucleon to be consistent with the well known principles. 

To assess the expression (\ref{resp}) we need to compare it with \eq{res1}, the most general expression for the nucleon current, $<P_+,s'|J^\alpha|P_-,s>$, in the relativistic 
quark-spectator diquark model (with diquark on mass shell) with 
the photon interacting only
with the quark (using a point-like photon-quark interaction and ignoring explicit
factors of charge), where $\bar\Gamma_\mu(P_+,k)$ and $\Gamma_\nu(P_-,k)$ are {\it covariant} 
functions of Dirac $\gamma$-matrices and the momentum variables. The factor 
$\frac{k^\mu k^\nu}{m^2}-g^{\mu\nu}$ arises from the axial-vector di-quark propagator and 
must
present in all quark spectator axial-vector di-quark models .

We stress that  
 (\ref{res1}) must hold independently of the choice of the particular form  
the diquark polarization vectors. A 
given  model is defined only by 
the choice of  a specific  form for 
$\bar\Gamma_\mu(P_+,k)$ and $\Gamma_\nu(P_-,k)$, the rest is fixed by  
(\ref{res1}), there is nothing left to  choose. 
The results of  \cite{ramalho} or any other model {\it must} be 
consistent with \eq{res1}, they should correspond to particular choices of a {\it covariant} $\Gamma_\nu(P_\pm,k)$. 

It is useful to discuss
the result (\ref{res1}) in the operator formalism of the quantum field theory. We compute the
matrix element $<P_+|\bar q\gamma^\alpha q|P_->$ by inserting the sum over a complete set of states:
\be
{\cal M}^\alpha=<P_+,s'|\bar q\gamma^\alpha q|P_-,s>
\sim\sum_n<P_+,s'|\bar q|n>\gamma^\alpha <n|q|P_-,s>.
\ee
Truncating the sum to the term of the valence quark model, $\sum|n><n|\rightarrow\int\sum|k,i><k,i|$, (where $|k,i>$ are diquark states) gives: 
\be
{\cal M}^\alpha\sim
\int d^4k\delta^+(k^2-m^2)\sum_i\bar 
U(P_+,s')\bar\Gamma_\mu(P_+,k)\gamma^\alpha \epsilon^\mu(k,i)
\bar\epsilon^\nu(k,i)\Gamma_\nu(P_-,k)U(P_-,s)\label{gen1}
\ee
where the diquark "conventional" polarization 
vectors, $\epsilon^\mu(k,i)$,
$\bar\epsilon^\nu(k,i)$, are exposed. The integrand of \eq{gen1} is a sum of contributions factorized in the 
nucleon momenta,  $P_+,P_-$, it is a sum of 
 terms each of which is of the 
form of 
product of scalar functions:$F_1(P_+,k)F_2(P_-,k)$.
 This property that the integrand 
is a sum of products is a signature property of a valence quark model (covariant and non-covariant).

Thus the relevant requirements for obtaining a correct evaluation of a
relativistic valence quark spectator axial di-quark model are summarized in
\eq{res1} and \eq{gen1}.

We  
find  that $D^{\mu\nu}(P_+,P_-)$ in \eq{resp} is indeed the covariant tensor given in (28) of \cite{ramalho}:
\bea
&&D^{\mu\nu}(P_+,P_-)=\sum\xi(P_+) \times \xi^T(P_-)=\Lambda\epsilon(Z_+) \times \epsilon^T(Z_-)\Lambda^T 
\nn
&&=\Lambda\left\{(Z^+_\mu +Z^-_\mu)(Z^+_\nu +Z^-_\nu)\frac{1}{M^2+Z_+\cdot Z_-}
-g_{\mu\nu}
-\frac{Z^+_\mu Z^-_\nu}{M^2}\right\}\Lambda^T 
\nn
&&=(P^+_\mu +P^-_\mu)(P^+_\nu +P^-_\nu)\frac{1}{M^2+P_+\cdot P_-}
-g_{\mu\nu}
-\frac{P^+_\mu P^-_\nu}{M^2}
\eea
But
a brief inspection shows that 
 the  result (\ref{resp})  is not 
consistent with the {\it general} form (\ref{res1}). 
The integrand  of  the general requirement
(\ref{res1}) is of the form of 
  a sum factorized of terms of the form  $F_1(P_+,k)F_2(P_-,k)$
in the nucleon momenta,  $P_+,P_-$, the integrand of the expression \eq{resp}
is  not factorisable due to the presence of the denominator 
$M^2+P_\cdot P_-$ in $D^{\mu\nu}(P_+,P_-)$.  There is no way 
to derive the integrand of (\ref{resp}) from (\ref{res1}).

We explain this in more detail by explicitly using the model  
 of \cite{ramalho} in \eq{resp}. This is to illustrate that the Dirac operators
do not yield a factor of ${M^2+P_+\cdot P_-}$ in the numerator that 
cancels the one in the denominator.  The model  takes the form $\Gamma'_\mu(P_-,k) =\gamma_u\gamma^5 \psi_1(P_-,k)$, with $\psi_1$ a scalar wave function.
Use these $\Gamma'_\mu$s and \eq{app} in \eq{resp} to obtain:
\bea\label{resp1}
{\cal M}^\alpha&=&
\int d^4k\delta^+(k^2-m^2)\psi_1(P_+,k)\psi_1(P_-,k)\bar U(P_+,s')\gamma_\mu D^{\mu\nu}(P_+,P_-)
\gamma^\alpha\gamma_\nu U(P_-,s)
\nn
&=&
\int d^4k\delta^+(k^2-m^2)\psi_1(P_+,k)\psi_1(P_-,k)\bar U(P_+,s')
\left[\frac{4M\;(P_+^\alpha+P_-^\alpha)}{M^2+P_+\cdot P_-}
-\gamma^\alpha\right]
 U(P_-,s).
\eea
The explicit appearance of a term 
 inversely proportional to $M^2+P_+\cdot P_-$  shows
that the model \cite{ramalho} violates the general requirements expressed in
Sects~\ref{sec:cov} and \ref{sec:uncon}. The use of the covariant expression (\ref{app}) for $D^{\mu\nu}$ leads
to a contradiction with well-known principles.

  It is important to see that 
  the transformation $\Lambda$ and vectors $Z_+$ and $Z_-$ depend on both momenta, $P_+$ and $P_-$ and thereby so do the unconventional polarization vectors (therefore the notation $\xi(P_\pm)$ with only one momentum argument is misleading)! This fact is not emphasized in 
  \cite{ramalho}. 
It might be interesting to show that this dependence on two momenta is completely contained in Wigner rotations,
$W_j^i(P_\pm,Z_\pm)$, usually involved in the Lorentz transformation of polarization vectors
  \be
\xi(P_+,i) =\Lambda\epsilon^i(Z_+)=\sum_j\epsilon^j(\Lambda Z_+)W_j^i(\Lambda Z_+,Z_+)=
\sum_j\epsilon^j(P_+)W_j^i(P_+,Z_+)
\ee
where  the polarization vector, $\epsilon^j(P_+)={\cal L}_{P_+}\epsilon^j_0$, depends only on $P_+$ just in the same way as the genuine polarization vector $\epsilon^j_k$ from (\ref{psi2}) depends on $k$. The whole dependence on the two momenta, $P_+$ and $P_-$, is contained in the Wigner (purely 3D) rotations, 
$W_j^i(P_+,Z_+)$,  through $Z_+=\Lambda^{-1} P_+$.
 
As a result the wave function of the initial (final) nucleon in the \eq{resp},
\be
\Psi_{P_-}\sim\bar\xi^\nu(P_-,i)\gamma^\alpha\Gamma'_\nu(P_-,k)U(P_-),\hspace{1cm}
\bar\Psi_{P_+}\sim\bar U(P_+)\bar\Gamma'_\mu(P_+,k)\xi^\mu(P_+,i), \eqn{gen-r-wf}
\ee
 depends not only on the momentum of the initial $P_-$ (final $P_+$) nucleon, it depends on both, $P_+$ and $P_-$, which is beyond the very idea of quark models. This dependence on the both momenta comes via the ''unconventional" diquark polarization vectors, they depend on both the initial and the final nucleon momentum, $P_+,P_-$.
 
 Having shown that the wave function of \cite{ramalho} corresponding to \eq{psiRel} is not consistent with  
well-known principles, here we can discuss restrictions these principles impose on the general form of $\bar\Gamma'_\mu(P_+,k)$ if one is to use the wave functions (\ref{gen-r-wf}).
 
The minimal consistency condition is
\bea
&&\sum_i\bar 
U(P_+,s')\bar\Gamma_\mu(P_+,k)\gamma^\alpha \epsilon^\mu(k,i)
\bar\epsilon^\nu(k,i)\Gamma_\nu(P_-,k)U(P_-,s)
\nn
&&=\sum\bar U(P_+s')\bar\Gamma'_\mu(P_+,k)
\xi^\mu(P_+,i)\bar\xi^\nu(P_-,i)\gamma^\alpha\Gamma'_\nu(P_-,k)U(P_-s)
\eea
which restricts $\Gamma'_\nu$ because $\Gamma_\nu(P_-,k)$ is restricted to being covariant and independent of $P_+$. This leads to 
\bea\label{sum-rest}
&&
\bar U(P_+,s')\bar\Gamma_\mu(P_+,k)\gamma^\alpha(\frac{k^\mu k^\nu}{m^2}-g^{\mu\nu})
\Gamma_\nu(P_-,k)U(P_-,s)
\nn
&&=\bar U(P_+s')\bar\Gamma'_\mu(P_+,k)
D^{\mu\nu}(P_+,P_-)\gamma^\alpha\Gamma'_\nu(P_-,k)U(P_-s).
\eea
Although \eq{sum-rest} admits a covariant solution for $\Gamma'_\nu(P_-,k) $ it requires $\Gamma'_\nu(P_-,k) $ to depend also  on $P_+$. A solution to \eq{sum-rest} is  
\be
\Gamma'_\nu(P_-,k)=({\cal L}_k R{\cal L}^{-1}_{Z_-}\Lambda^{-1})^\beta_\nu
\Gamma_\beta(P_-,k)
,\hspace{1cm}
\bar\Gamma'_\mu(P_+,k) =\bar\Gamma_\alpha(P_+,k) ({\cal L}_k R{\cal L}^{-1}_{Z_+}\Lambda^{-1})^\alpha_\mu,
\eqn{P+P-}
\ee 
where $R$ is an arbitrary 3D rotation. The mentioned dependence on the momenta of the both nucleons comes from
${\cal L}_k R{\cal L}^{-1}_{Z_+}\Lambda^{-1}$. The solution (\ref{P+P-}) can be verified by using the identity
\be
\frac{k^\alpha k^\beta}{m^2}-g^{\alpha\beta}
=({\cal L}_k R{\cal L}^{-1}_{Z_+}\Lambda^{-1})^\alpha_\mu D^{\mu\nu}(P_+,P_-)({\cal L}_k R{\cal L}^{-1}_{Z_-}\Lambda^{-1})^\beta_\nu
\ee
The arbitrariness  of $R$ follows from the identity
 \be
R^\alpha_\mu (\delta^{0\mu} \delta^{0\nu}-g^{\mu\nu})R^\beta_\nu=\delta^{0\alpha} \delta^{0\beta}-g^{\alpha\beta}
\ee
 
 The expressions in \eq{sum-rest} are closely related to the generalized parton distributions 
(GPD) corresponding to the parton (quark) momentum $P_--k$. We would like to note that for 
any given model of the nucleon one can construct such GPD related function\cite{gpd} which reproduces
 exactly the em current of the given model and at the same time it corresponds to the round 
spin-dependent matter density. The problem is that in such GPD the initial and final nucleon
 momenta are not factorized and therefore it cannot be explained in the framework of a valence quark model.
\section{Assessment}
\label{sec:assess}
Using the first of two interpretations discussed above (conventional polarization vectors), 
we have shown that
the seemingly covariant appearance of the expressions of \cite{ramalho}
results from the explicit use of the  Breit frame. This failure
to maintain covariance results from using 
 the polarization vectors $\varepsilon_P$ instead of $\epsilon_k$ to describe
the vector di-quark wave function. 
However, this is a very important point in the present context because it is
exactly the use of  $\varepsilon_P$ that allows 
the construction of a model wave function without orbital angular momentum.
 As noted in 
Ref.~\cite{ramalho}, the result (\ref{app}) has no angular dependence,
so  the evaluation of the matrix element of the 
 spin-dependent density operator would yield  a spherical shape. However, this
roundness is caused 
{solely} 
by 
the lack of Lorentz invariance. 
Using the polarization  vector $\epsilon_k$ would lead to a model
much like that of \cite{Gross06}, which 
does have   a non-spherical shape, as measured by the
spin-dependent matter density. 

Given the importance of 
 Lorentz invariance, we have derived the general form of the wave function 
that  would produce a Lorentz invariant em form factor. This
 derivation led us unambiguously to the 
''unconventional" diquark polarization vectors $\xi(P_\pm)$,  suggested in \cite{Franz},
 to interpret the wave functions used in \cite{ramalho}.  
 These  yield covariant results.
Unfortunately neither the derived wave functions 
nor the expression for the em current the satisfy well-known 
principles discussed in the Sect. \ref{sec:uncon}. For example, 
the wave function of the incoming nucleon depends on the momentum of the outgoing nucleon too and vice versa, so 
the em current 
cannot be written in the traditional form of the convolution of two proper wave functions.

\section*{Acknowledgments}
We thank the USDOE and the CRDF grant 
 GEP2-3329-TB-03  and the GNSF grant No GNSF/ST06/4-050  for partial support of this work.
We thank W.~Detmold for commenting on the manuscript. We  thank Prof. Franz Gross for useful discussions and for supplying us with the
unpublished material in Ref.~[18].



\begin{thebibliography}{9}
\bibitem{diqe} We consider only the contributions of the vector-diquark component of the nucleon
wave function here.
\bibitem{explain}

Note that the boost (\ref{boost1}) differs from the traditional one, which is used
in  \cite{ramalho}. Our ${\cal L}^{-1}_k$
 coincides with $B^{-1}_k$ from \cite{ramalho} for ${\bf k}$ aligned with the $z$-axis.
Ref.~\cite{ramalho}  uses $B^{-1}_kR^{-1}(k)={\cal L}^{-1}_zR^{-1}(k)$ 
instead of our ${\cal L}^{-1}_k$ where
$z=(k^0,0_\perp, |k|)$ and  $R$ is the 3D rotation ($R^{-1}(k)k=z$). The 
difference between our boost and that of \cite{ramalho} is a
 matter of only tradition, because both are valid boosts. 
If we used the boost of  \cite{ramalho}  the function
$D_{\mu\nu}$ obtained in \eq{Dmunu2} would be replaced by
\be
D_{\mu\nu}(P_+,P_-)\rightarrow R_\mu^\alpha(P_+)D_{\alpha\beta}(Z_+,Z_-)R_\nu^\beta(P_-),
\eqn{trad}
\ee
where $Z_\pm=(P_\pm^0,0_\perp,|P_\pm|)$. To evaluate \eq{trad}
we  use
\bea
R(P_\pm)Z_\pm&=& P_\pm
\nn
(R(P_+)Z_-)_\mu
&=&[R(P_+)(P_-^0,0_\perp,|P_-|)]_\mu=(P_-^0,\frac{|P_-|}{|P_+|}\vec{P_+})_\mu
\nn
&=&
\frac{|P_-|}{|P_+|}P_\mu^++(P_-^0-\frac{|P_-|}{|P_+|}P_+^0)\delta^0_\mu
\eea
The result of this exercise would lead to a  $D_{\mu\nu}$ that
contains even more factors that violate covariance 
  than the  $D_{\mu\nu}$ obtained in \eq{Dmunu2}.
\end{thebibliography}
\end{document}